\documentstyle[twoside,fleqn,espcrc2,psfig]{article}

\newcommand{\ltsima} {$\; \buildrel < \over \sim \;$}
\newcommand{\gtsima} {$\; \buildrel > \over \sim \;$}
\newcommand{\lta} {\lower.5ex\hbox{\ltsima}}
\newcommand{\gta} {\lower.5ex\hbox{\gtsima}}

\newcommand{\nodata} {\raise.5ex\hbox{....}}

\newcommand{\AmS}{{\protect\the\textfont2
  A\kern-.1667em\lower.5ex\hbox{M}\kern-.125emS}}

\title{ A first look at blazars with {\it Beppo}SAX}

\author{
L. Maraschi\address{Osservatorio Astronomico di Brera, via Brera
28, Milano, Italy} 
}

\begin{document}

\begin{abstract}
I describe a general framework that could allow to understand the broad band
spectra  of blazars and lead to a unified picture of the emission from
relativistic jets, in BL Lac objects as well as in flat spectrum, radio loud Quasars. 
The scheme serves as a useful basis to introduce and discuss some of the most interesting
results so far obtained on Blazars with {\it Beppo}SAX. 
 \end{abstract}

\maketitle

\section{Introduction}

The "blazar phenomenon" is due to the presence of  relativistic flows
(jets) emanating from the nuclei of active galaxies which are
radio loud. The power to energize
the radio lobes is transported in the jets. 
Blazars are the subset of radio loud AGN  for which the  relativistic jet
happens to point at small angles to the line of sight. Since the 
radiation emitted by the jet is relativistically beamed into a narrow cone 
along the
direction of motion,  the aligned observer will receive a strongly
enhanced flux. For a bulk Lorentz factor $\Gamma\simeq 10$ at an angle
$\theta\simeq 1/\Gamma$ the flux enhancement factor is $10^3-10^4$.

The evidence in favor of this picture has been accumulating and is now
solid (e.g. \cite{pu} and refs therein)
 although the origin of the jets is poorly 
understood and their physical parameters are highly uncertain.
The relativistically amplified,  non thermal emission from high energy
particles in the jet can account for the extreme properties of blazars 
concerning variability, polarization and energy distribution of the continuum, 
which extends from the radio to the gamma-ray band. 

Traditionally BL Lac objects, where no prominent emission lines are observed
(with an upper limit of 5 \AA), were thought to represent a separate class
perhaps more extreme than Quasar-like blazars. The latter include
optically violently variable
and highly polarized quasars (OVV, HPQ) or more generally Quasars with flat radio
spectrum (FSRQ) indicating strong emission from the self absorbed core.
It has become clear however that BL Lacs have on average lower luminosity
than quasar like blazars (\cite{pado}) and that the distribution of
emission line equivalent widths is  continuous (\cite{scafal}).
We will therefore in the following consider blazars as a single class of
objects, implicitly assuming that, irrespective of the emission line properties
which derive from the surrounding gas, the same physical mechanisms operate
in  relativistic jets over a wide range of luminosities.

By studying the blazar continuum we expect to learn about the radiation mechanisms
in the jets, about the processes of particle acceleration and energy transport
along the jets and ultimately about their origin and evolution.   

\begin{figure*}[bt]
\vspace{9pt}
\psfig{file=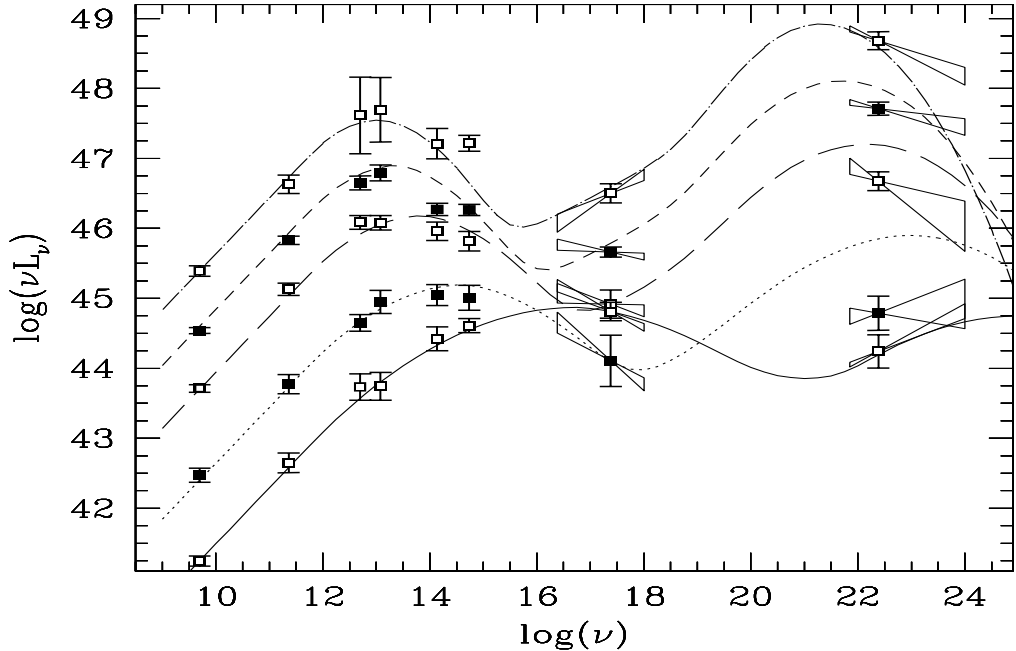,width=15.0truecm,height=11.5truecm,rheight=8.7truecm}
\caption{\small\sf Average SEDs for the ``total blazar sample'' binned 
according to radio luminosity irrespective of the original
classification.  The overlayed dashed curves are analytic approximations
obtained assuming that the ratio of the peak frequencies is constant and that the
luminosity at the second peak is proportional to the radio luminosity (from
\protect\cite{foss98} }
\label{fig:sed_medie}
\end{figure*}

\section {The broad band spectra of blazars}
 
The discovery by EGRET on board CGRO of copious $\gamma$-ray emission from blazars
caused a "Renaissance" in this field. In fact it had been noted 
early on that the high
density of relativistic electrons necessary to produce the observed compact
synchrotron emission would lead to strong, even catastrophic inverse Compton
radiation (\cite{hbs}).  X-ray measurements were used to constrain the
amount of inverse Compton emission allowed and to derive minimum values
for the necessary beaming factors (e.g. \cite{gp93}). At present the
 $\gamma$-ray observations allow to
measure the intensity and spectral shape of a component which contains a
substantial fraction and in some cases the bulk of the emitted power
leading to
strong constraints on the physical parameters of the emitting region.

It is clearly important, besides observing single objects, to try to derive 
general properties of the continuum and understand whether  and how they differ
for instance in BL Lac objects and FSRQ. It is especially interesting to
discuss whether the gamma-ray emission is a general property of the whole class.
 We have recently addressed this problem (\cite{foss98})
collecting multifrequency data for three complete samples of blazars:
the 2 Jy sample of FSRQs, the 1Jy sample of BL Lac objects and the sample of BL Lacs
selected in the X-ray band from the Einstein Slew Survey.
Systematic differences in the shape of the continuum in specific spectral bands
 among different subclasses 
of blazars were noted early on (e.g. \cite{gg86,imp,ww,smu,umu}.
In particular we note that the percentage of objects detected with EGRET
(100 MeV - 10 GeV) is significantly larger for the sample of FSRQ than for the two 
BL Lac samples (40 \% vs. 26\% and
17\% respectively. Nevertheless plotting the average SEDs as shown in 
Fig.~\ref{fig:sed_medie}, we can see that the shapes are "globally similar".

 In Fig.~\ref{fig:sed_medie} all blazars in
the three complete samples were merged and  grouped in luminosity classes 
irrespective of their
original classification and the dashed lines drawn for comparison derive from
an analytic parametric representation (\cite{foss98}). 

The main results of this work are the following:
\begin{itemize}
\item two peaks are present in all the SEDs

\item the first peak occurs at lower frequencies for the highest luminosity objects

\item the frequency at which the second peak occurs correlates with that of the
 first one. The dashed curves correspond to a constant ratio between the two 
peak frequencies.             

\end{itemize}

For the most luminous objects the first peak is at frequencies lower than
the optical band while for the least luminous ones the reverse is true.
Thus highly luminous objects have a "red" (steep) IR to UV continuum while 
objects of lower luminosity have a bluer IR to UV continuum. For this reason
and to recall intuitively the location of the peaks on the
frequency axis we will briefly call 
"red" blazars the objects in the three highest luminosity classes and 
"blue" blazars those in the two lower luminosity classes. The present data
suggest a continuous spectral sequence and no absolute separation between 
red and blue blazars.

Considering the continuum of different objects in a fixed spectral range
 its shape changes systematically with luminosity along the sequence,
as the peak frequency approaches and  moves across the chosen frequency
interval. In particular the X-ray spectrum becomes steeper and the gamma-ray
spectrum (in the EGRET range) becomes flatter from "red" to "blue" blazars,
as the two peaks march to higher
frequencies. The different location of the gamma-ray peak can account for
 the different detection rates of BL Lacs and FSRQs by EGRET. Objects 
whose $\gamma$-ray emission peaks in the EGRET range are more easily detected.

Recently ground based observations in the TeV range performed with Cherenkov telescopes
have detected two of the X-ray brightest "blue" blazars (refs). We expect that with the 
progress in sensitivity  many more will be detected giving access to the study
of the highest energies from ground.

A final comment concerns variability. It is interesting to note that 
the largest variability is usually observed close to or above each of the two 
peaks and is usually in the sense of a hardening of the spectrum at higher
intensity. These statements are  based mostly on observations at high
energies (X-rays and gamma-rays) and concern a limited number of sources
(e.g. \cite{umu}) therefore they should be considered as 
tentative suggestions rather than established facts. 

\section {Interpretation}

It is generally thought that the first spectral component peaking at far
infrared up to X-ray frequencies is due to synchrotron radiation.
The spectra from the radio to the submm range most likely involve
contributions from different regions of the jet with different
self absorption turnovers. However, from infrared frequencies upwards 
the synchrotron emission should be thin and could be produced 
essentially in a single homogeneous region. 
Inverse Compton scattering of soft photons by the high energy electrons emitting
 the thin synchrotron radiation could be responsible for the second ( high frequency)
component of the SED, peaking in the gamma-ray band.  
The soft photons could be the synchrotron photons themselves (SSC) or
photons outside the jet (EC), possibly produced by an accretion disk or torus
 and scattered or reprocessed by the surrounding gas (e.g \cite{sbr}, \cite{umu} and 
refs therein).   

If the same region is responsible for the two spectral components then, 
irrespective of the nature of the seed photons, {\it the two peaks must derive
from  the same high energy electrons}. Therefore a change in the 
density and/or spectrum of those electrons is expected to cause
correlated variability at frequencies close to the two peaks.
In the SSC model the inverse Compton intensity is expected to vary more
than the synchrotron one, approximately as the square of it in the simplest
case while in the EC model one expects a linear relation (\cite{miami}).
 
Measuring the two peaks simultaneously is thus the best means to determine the
physical parameters of the emission region and studying the variability
of the spectra around the peaks yields unique insight into the mechanisms
of particle acceleration and  energy loss in the jet. 
The variability correlation should enable to disentangle the contribution of different
sources of seed photons (SSC vs. EC).

The "spectral sequence" discussed above could be attributed to a systematic
dependence of the critical electron energy (the break energy) and/or of the
magnetic field on luminosity. Assuming that the beaming factors are not
significantly different along the sequence, the trend in apparent luminosity
is also a trend in intrinsic luminosity. In the SSC model the break energy 
of the electrons is univocally determined by the ratio of the frequencies of the two
peaks and should therefore be approximately constant. "Red" blazars should then
have lower magnetic field than "blue"
blazars. Systematic model fitting of all the $\gamma$-ray detected blazars 
with sufficient multifrequency data suggest  that 
 as the magnetic  energy density decreases
the external photon energy density becomes important so that a smooth transition
between the SSC and EC scenario takes place (\cite{gg98}).

\section {SAX observations}

The X-ray band is crucial for a discussion of the above problems in that
the synchrotron and inverse Compton components which have different
spectral shapes may both be relevant. Simultaneous observations over 
a broad energy range are 
required to disentangle the two mechanisms. The  unique characteristics of the
{\it Beppo}SAX instrumentation  appear therefore ideal for blazar studies.
  Observations of bright blazars detected in $\gamma$--rays were
proposed with the main aims of:

\begin{itemize}
\item determining the spectral shape up to the 100
keV range, thus exploring the connection between X-rays and gamma-rays

\item studying the variability 
in relation with other wavebands, especially $\gamma$--rays.
\end{itemize}

\begin{figure*}[bt]
\vspace{9pt}
\psfig{file=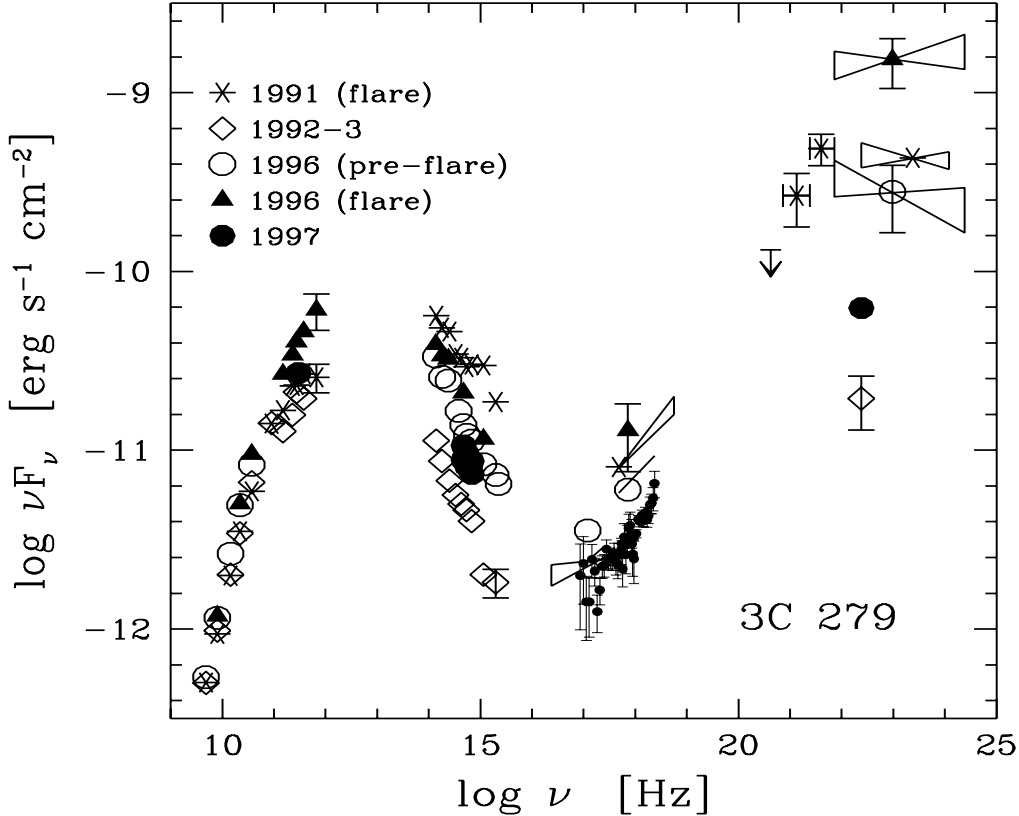,width=15.0truecm,height=13truecm,rheight=11.2truecm}
\caption{\small\sf Spectral energy distribution during the 1997 campaign,
compared with previous ones. The 1997 data are from: X--rays = 
{\it Beppo}SAX data; $\gamma$--rays = Hartman, private comm.; R-band =
Raiteri \& Villata, private comm.}
\label{fig:sed_3c279}
\end{figure*}

In the following I will briefly mention and comment some of the most interesting
results obtained so far. Besides 3C 273 (\cite{htv}) which is
probably intermediate between a blazar and a "normal" quasar,  two flat spectrum 
radio quasars, 3C 279 and PKS 0528+134, were observed (before June 1997)
and found in a low 
intensity state.  In the scheme presented above these are "red" blazars.
 I will discuss here the first source (also \cite{mtv}), while for the second one 
I refer to \cite{geltv}. 
 Finally I will consider results on "blue" blazars ( Mrk 421, 1ES 2344+514 and
Mrk 501 (see the contributions in this volume \cite{ftv}, \cite{gitv} and  
\cite{gtv} respectively).  
 
It is important to remember that in "red" blazars the X-ray emission represents
the lower energy end of the inverse Compton emission, while for "blue"
blazars it is the high energy end of the synchrotron emission.

\subsection {3C 279}

The X-ray spectrum measured with {\it Beppo}SAX in January 1997, the
simultaneously measured optical flux  and the quasi simultaneous gamma-ray
flux (Hartman, private communication) are shown in Fig.~\ref{fig:sed_3c279} 
together with
other simultaneously measured SEDs obtained at other epochs: the high
state observed in June 1991 (\cite{hart}), the low state observed in
January 1993 (\cite{m94}) and the preflare and flare states observed in
January - February 1996 (\cite{wehrle98}).

At the epoch of the {\it Beppo}SAX observations the  $\gamma$--ray
flux was a factor 6 and 20 weaker respectively than  measured in June 1991
and early February 1996.
In X-rays the amplitude is smaller but there is a good correlation between the X-ray
and gamma-ray fluxes especially in the 2-10 keV band (see Fig. 3 in Maraschi et al.
this volume). Note that the fluxes at 1 keV measured by
{\it ROSAT} and {\it Beppo}SAX  in the 1993 and 1997 low states 
 are similar   However
the spectrum measured by {\it Beppo}SAX  is significantly flatter providing a good
connection with the higher $\gamma$--ray flux in 1997.
 The simultaneity (within one day) of the X-ray (XTE) and  $\gamma$--ray peaks
during the 1996 flare suggests that the X-ray to gamma-ray emission originates in
a single region and that the spectrum hardens with increasing intensity. It
is possible  that the $\gamma$-ray peak in the SED  moves to higher energies
at the flare peak.

The situation  is much more complex at lower energies. Although there is still a
general correlation of the IR-optical-UV fluxes with the gamma-ray intensity  
on long timescales, the flux variation at optical wavelengths corresponding to the
rapid 1996 flare is quite small. Note also that for 3C 279 the (presumed) peak of the
 synchrotron component falls in an unexplored region of the spectrum, between $10^{12}$
and $10^{14}$ Hz.

In the SSC model one expects that the inverse Compton emission varies with the 
square of the amplitude of the synchrotron emission due to the same electrons
(e.g. at the two peaks of the SED). This is compatible with the long term variations but not with the
strong rapid flare observed in 1996, where the amplitude in gamma-rays
was larger than the square of the optical one.
On the other hand if the seed photons for the inverse Compton process are external 
to the jet, they should not be rapidly variable and the inverse Compton emission 
is expected to vary linearly with the synchrotron one. 
Thus neither of the two "simple" models can adequately account for the multifrequency
variability behaviour.

A possible way out is that the seed photons derive from backscattering and/or
reprocessing of radiation produced in the jet by gas clouds closely approaching the
jet itself (\cite{ggmadau}). This model is attractive and needs to be studied
in more detail. Another possible way out is that the region emitting the synchrotron
radiation is inhomogeneous so that the observed variability is diluted by a more
stationary component.
  
\subsection {Blue blazars}

The X-ray emission from these objects has been observed to vary dramatically
on short timescales at least in the brightest prototypes, PKS 2155-304 and
Mrk 421. This can be understood recalling that
 the X-ray emission represents the high energy end of                 
 the synchrotron component: it is therefore due to radiation from the highest
energy electrons which have the shortest lifetimes and can vary very rapidly.   

A continuous acceleration mechanism must be responsible for maintaining
in the source particles which have lifetimes of the order of hours.
However the injection mechanism may be continuous only in an average sense
or at a low intensity level and  episodes of increased injection rate
may occur causing variability. The study of flares and of spectral variability
associated with them  gives direct information on the spectra of the
freshly injected/accelerated particles and their subsequent decay
to a state of quasi-equilibrium.  
 
The photons upscattered through the inverse Compton process by the highest
energy electrons reach  TeV energies so that the "bluest" and
brightest sources can be detected from ground based Cherenkov telescopes.
 This is the case up to
now for three objects: Mrk 421, 1ES 2344+514 and Mrk 501.
All of them were observed with SAX in the first half of AO1.
The simultaneous observation of a source in X-rays and at TeV
energies should allow to determine unambiguously the energy of the radiating
electrons, the beaming factor, the magnetic field and the energy
density of the seed photons. In some cases it may be necessary to take into
account that scattering will occur in the Klein Nishina regime.

\begin{figure*}[bt]
\vspace{9pt}
\psfig{file=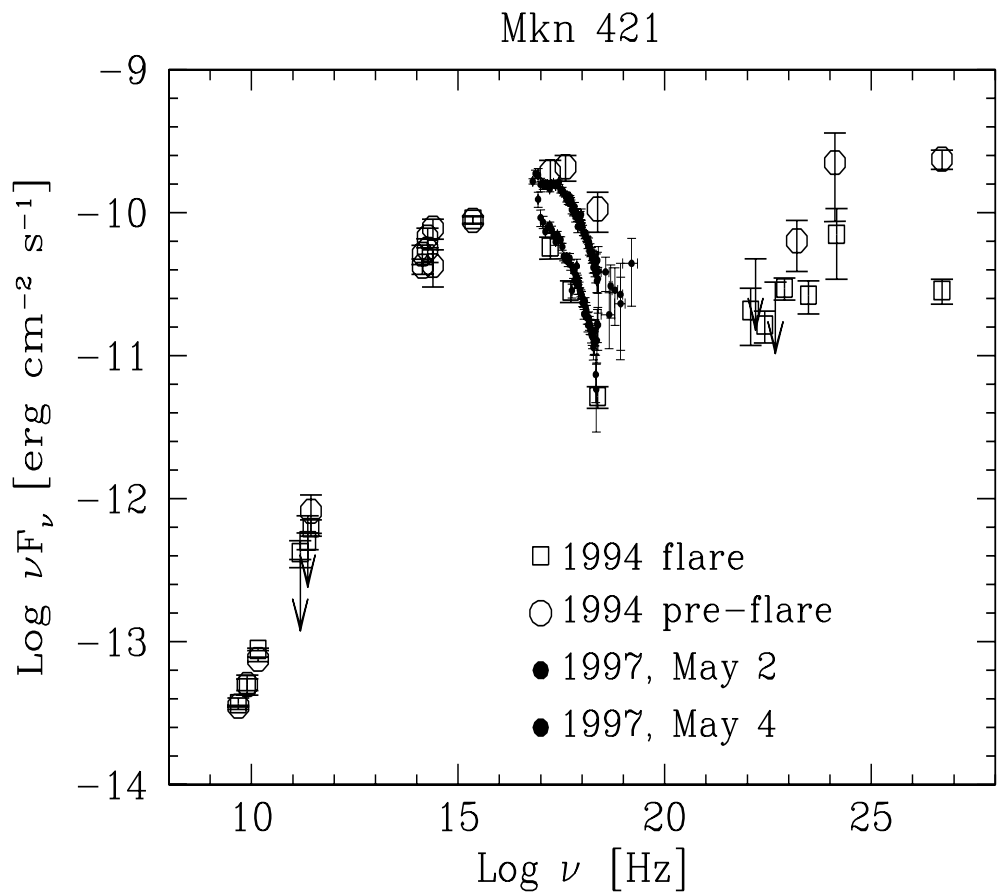,width=15.0truecm,angle=270,height=13truecm,rheight=11.2truecm}
\caption{\small\sf The May 97 data from {\it Beppo}SAX are compared
with the  energy distributions measured in 1994 (\protect\cite{macomb}}
\label{fig:sed_mkn421}
\end{figure*}

\subsection {Mrk 421}

This source has been repeatedly observed with ASCA and at other
relevant wavelengths including TeV observations. (\cite{macomb}) 

The {\it Beppo}SAX observations of Mrk 421 (Fossati et al. this volume)
show a decay between 
two intensity states closely similar to those  previously observed with ASCA.
The higher state shows a flatter spectrum indicating that the emission peaks
 at higher energies. In Fig.~\ref{fig:sed_mkn421}
 the {\it Beppo}SAX data are compared with the
ASCA and TeV data presented by \cite{macomb}.

Mastichiadis and Kirk \cite{mk}have shown that this behaviour could be 
the result of an injection mechanism in which the maximum energy of the
injected particles increases. The two "states"
would therefore represent equilibrium states for injection spectra extending
to different maximum energies.   
The fact that the spectral variability observed in 1994 is closely reproduced
in the  {\it Beppo}SAX observations
indicates that the involved region is the same (same physical 
parameters). 
In addition SAX detected the source at higher energy with the PDS. The
preliminary analysis yields a flat spectrum at high energies 
suggesting that the inverse Compton component becomes dominant in this 
band. The variation in the PDS is much smaller than in the MECS which
is consistent with attributing the hard emission to IC from 
electrons of much lower energies.
An extensive campaign for simultaneous ASCA and TeV observations will take place
in  the spring of 1998, to which {\it Beppo}SAX observations could add significantly.

\subsection {1ES 2344+514, Mrk 501}

This source is not as well studied as Mrk 421 but  shows 
a  similar, more extreme behaviour (\cite{gitv}).
 The emission peak was in the medium X-ray range in December 96 and
shifted to 20-50 keV during the flare of a factor 2 observed between December 3-7 96.

The most extraordinary spectral variation was found in Mrk 501
(\cite{pian98}, \cite{gtv}). The source was in a 
state of strong activity at TeV energies although the 1 keV flux
had not increased dramatically. However compared to previous
 observations {\it the X-ray spectrum had changed dramatically}
being flatter than or close to 1 up to 100 keV. In correspondence
to a large TeV flare , the 1 to 100 keV spectrum hardened still 
indicating that the emission peak was at or above 100 keV, while 
past multifrequency measurements  showed it to be below 1 keV.
Thus in Mrk 501 the shift of the peak frequency was more than a 
factor 100. Some theoretical implications are discussed by Ghisellini
(\cite{gtv}).

This unprecedented behaviour may be less uncommon than judged at first
sight. In fact in the medium X-ray band the intensity behaviour
was not exceedingly dramatic and good spectral capabilities 
up to the hard X-ray band, necessary to reveal the phenomenon,
have only recently become available with {\it Beppo}SAX.

It is interesting to mention that a spectral survey of BL Lacs
selected in relatively soft X-rays yields evidence of some 
hard X-ray spectra (\cite{wolter}). Sources which have 
more or less permanent emission peaks in the hard X-ray range may also exist
and have gone undetected so far.

\section{Conclusions}

 The results  from the first part of the {\it Beppo}SAX  AO1 Core Program 
on Blazars are extremely exciting. Bright $\gamma$-ray blazars can be
usually detected up to 50 - 100 keV allowing detailed study of the
synchrotron and inverse Compton emission and of the correlations
between their temporal and spectral variability.

The observations  presented above suggest and support:   

\begin{itemize}
\item the correlation between the X--ray (1 keV)
and $\gamma$--ray (0.1-10 GeV) fluxes, which now holds over a period
of 6 years and about a factor 30 in flux change for $\gamma$-rays
(3C 279 -- PKS 0528 + 154)

\item the smaller amplitude of variability of the synchrotron
component compared to the high energy one. 

\item the synchrotron emission peak seems to shift systematically to higher
energies during flares (Mrk 421, 1ES 2344+541, Mrk 501)

\end{itemize}

These results will undoubtedly stimulate new observations and
new theoretical approaches.  In particular genuinely time dependent
 models  for the acceleration of particles and the spectral evolution of   
the emitted radiation are needed.

\end{document}